\documentclass{emulateapj}
\usepackage{graphicx}
\usepackage{natbib}
\usepackage[usenames,dvips]{color}
\usepackage{amssymb, amsmath, amsbsy, epsfig, epsf}
\usepackage{booktabs}
\usepackage{longtable}
\def\hb{H$\beta$}
\def\ha{H$\alpha$}
\def\feii{\ion{Fe}{2}}
\def\mgii{\ion{Mg}{2}}
\def\civ{\ion{C}{4}}
\def\oiii{[\ion{O}{3}]}
\def\nii{[\ion{N}{2}]}
\def\sii{[\ion{S}{2}]}

\def\kmps{${\rm km\,s^{-1}}$}

\slugcomment{}
\shorttitle{}
\shortauthors{}

\begin{document}

\title{The Large Sky Area Multi-Object Fiber Spectroscopic Telescope Quasar Survey: Quasar Properties from First Data Release}

\author{
Y.L.Ai\altaffilmark{1,2},
Xue-Bing Wu\altaffilmark{1,3},
Jinyi Yang\altaffilmark{1,3},
Qian Yang\altaffilmark{1,3},
Feige Wang\altaffilmark{1,3},
Rui Guo\altaffilmark{1,3},
Wenwen Zuo\altaffilmark{4},
Xiaoyi Dong\altaffilmark{1}, 
Y.-X. Zhang\altaffilmark{5},
H.-L. Yuan\altaffilmark{5}, 
Y.-H. Song\altaffilmark{5},
Jianguo Wang\altaffilmark{6,7},
Xiaobo Dong\altaffilmark{6,7},
M. Yang\altaffilmark{5},
H.-Wu\altaffilmark{5},
S.-Y. Shen\altaffilmark{4},
J.-R. Shi\altaffilmark{5},
B.-L. He\altaffilmark{5},
Y.-J. Lei\altaffilmark{5},
Y.-B. Li\altaffilmark{5},
A.-L. Luo\altaffilmark{5},
Y.-H. Zhao\altaffilmark{5},
H.-T. Zhang\altaffilmark{5}}

\altaffiltext{1}{Department of Astronomy, School of Physics, Peking University, Beijing 100871, China, aiyl@pku.edu.cn}
\altaffiltext{2}{School of Physics and Astronomy, Sun Yat-Sen University, Guangzhou 510275, China}
\altaffiltext{3}{Kavli Institute for Astronomy and Astrophysics, Peking University, Beijing 100871, China}
\altaffiltext{4}{Shanghai Astronomical Observatory, Chinese Academy of Sciences, Shanghai 200030, China}
\altaffiltext{5}{Key Laboratory of Optical Astronomy, National Astronomical Observatories, Chinese Academy of Sciences 100012, Beijing, China}
\altaffiltext{6}{Yunnan Observatories, Chinese Academy of Sciences, Kunming 650011, China}
\altaffiltext{7}{Key Laboratory for the Structure and Evolution of Celestial Objects, Chinese Academy of Sciences, Kunming 650011, China}
\email{aiyl@pku.edu.cn}

\begin{abstract}
We present preliminary results of the quasar survey in Large Sky Area Multi-Object Fiber Spectroscopic Telescope (LAMOST) first data release (DR1), which includes pilot survey and the first year regular survey.  
There are 3921 quasars identified with reliability, among which 1180 are new quasars discovered in the survey. These quasars are at low to median redshifts, with highest $z$ of 4.83.
We compile emission line measurements around the \ha, \hb, \mgii, and \civ\ regions for the new quasars.  The continuum luminosities  are inferred from SDSS photometric data with model fitting as the spectra in DR1 are non-flux-calibrated. We also compile the virial black hole mass estimates, and flags indicating the selection methods, broad absorption line quasars. The catalog and spectra for these quasars are available online. 28\% of the 3921 quasars are selected with optical-infrared colours independently, indicating that the method is quite promising in completeness of quasar survey.  LAMOST DR1 and the on-going quasar survey will provide valuable data in the studies of quasars.

\end{abstract}


\section{Introduction}
A huge effort has been undertaken to find quasars since the first one was discovered in 1963 (Schmidt 1963).
The most distant quasar, z=7.085, was identified by United Kingdom Infrared Telescope (UKIRT) Infrared Deep Sky Survey (UKIDSS) (Mortlock et al. 2011). Quasars, with accretion onto their central supermassive black holes, emit  a broad range of electromagnetic waves, from radio to $\gamma$-ray (Antonucci 1993). They can be one hundred times brighter than their host galaxies. Quasars are important extragalactic objects in astrophysics, which are used to map black hole growth (e.g.,Kollmeier et al. 2006), probe the galaxy evolution and intergalactic medium (e.g., Hennawi \& Prochaska 2007; Ross et al. 2009). 

Quasars can be selected from stellar and normal galaxies in different wavelength ranges, from radio to $\gamma$-ray, based on their distinguished spectral energy distribution, huge luminosity and variations.
The most yields of quasars are from Sloan Digital Sky Survey (SDSS)(e.g., Paris et al. 2012; Shen et al. 2011). Two-Degree
Fields survey (2dF) has also identified a large sample of quasars (Boyle et al. 2000). Both SDSS and 2dF select quasar candidates mainly based on multi-colour ultraviolet (UV)/optical photometric data, i.e., UV excess (Fan et al. 2000; Richards et al. 2002, 2004, 2009;  Smith et al. 2005 ). The completeness and efficiency of this selection are higher at redshifts less than 2, but drop significantly at $2.2\leq z \leq 3.5$, in which range is the peak of quasar luminosity function (Fan 1999; Richards et al. 2002, 2006; Schneider et al.
2007). The efficiency can be improved with combining of quasar selection from variability (Palanque-Delabrouille et al.. 2011; MacLeod et al. 2011; Graham et al. 2011; Morganson et al. 2014), of which multi-epoch data are required.

Infrared survey is promising in the potential of finding quasars at high redshifts and improving the completeness and efficiency of quasar survey at low and median redshifts. More quasars are identified with the UKIDSS survey (e.g., Warren et al. 2000; Croom et al. 2001; Sharp et al. 2002; Hewett et al. 2006; Chiu et al. 2007; Maddox et al. 2008, 2012; Smail et al. 2008), and studies are explored to select quasar candidates with SDSS/UKIDSS colours (e.g., Wu \& Jia 2010; Wu \& Jia 2011). Compared to the limited survey area of UKIDSS, the all sky survey Wide-field Infrared Survey Explorer (WISE)  (Wright et al. 2010, Cutri et al. 2012) data release shines more light in quasar survey.  Wu et al. (2012) proposed criteria for quasar candidate selections with SDSS/WISE colours. With the method  they detected an ultra-luminous  quasar at redshift of 6.3  (Wu et al. 2015).

Recently data-mining algorithms have been applied to the search for quasar candidates. The methods have been exploited  in the SDSS-III Baryon Oscillation Spectroscopic Survey (BOSS) (Ross et al. 2012), including an extreme-deconvolution method (XDQSO, Bovy et al. 2011), a kernel density estimator (KDE, Richards et al. 2004, 2009), a likelihood method (Kirkpatrick et al. 2011), and a neutral network method (Y\'eche et al. 2010). Peng et al. (2012) also develop a classification system based on support vector machine (SVM). Generally different selection methods are combined to reach a  higher completeness in quasar sky survey.

Large Sky Area Multi-Object Fiber Spectroscopic Telescope (LAMOST) is the Chinese multi-fiber telescope. Some quasar candidates, selected by optical-infrared colours, have been identified during the commission survey phase (Wu et al. 2010). LAMOST regular survey began in 2012, and there are $\sim$ 70,000 quasar candidates observed before 2013 June, as released in the LAMOST first data release. These objects are selected with several quasar selection methods. In this paper we describe LAMOST quasar survey and candidate selections in Section 2. Spectral measurements and black hole mass estimations for the new quasars are described in Section 3 and Section 4, respectively.  Section 5 presents the description of quasar catalog.  Section 6 is the discussion. We use a $\Lambda$-dominated cosmology with $h = 0.7$, $\Omega_{m} = 0.3$, and $\Omega_{\Lambda} = 0.7$ throughout the paper.

\section{Survey Outline}
LAMOST is a reflecting Schmidt telescope with an effective aperture of 4 m. It is equipped with 4000 fibers that can be deployed across 5$\degr$ (diameter) field of view. The spectral resolution is R $\sim$ 1800 over the wavelength range from 3800 \AA\ to 9100 \AA, which is divided into red and blue spectroscopic channels. Two arms of each spectrograph cover this wavelength range with an overlap of 200 \AA. The blue spectral coverage is 3700 \AA\ $\sim$ 5900 \AA, and the red is 5700 \AA\ $\sim$ 9000 \AA\ (Cui et al. 2012; Zhao et al. 2012). The raw data were reduced with LAMOST two-dimensional pipelines, including bias subtraction, cosmic-ray removal, spectral tracing and extraction, wavelength calibration, flat fielding, and sky subtraction (Luo et al. 2012). 

LAMOST began pilot survey in October 2011 and regular surrey in September 2012. The regular survey, carried out in five to six years, has two major components: the LAMOST Experiment for Galactic Understanding and Exploration (LEGUE) and the LAMOST Extragalactic Survey (LEGAS)(Zhao et al. 2012). LEGAS  only uses a small part of  available observing time due to the limitation of LAMOST site, especially the bright sky background and poor seeing. 
The first data release (DR1)  contains  spectra  taken before June 2013 (Luo et al. 2015). In this paper we present the results of quasar survey from  LEGAS, while the quasars discovered in the vicinity of the Andromeda and Triangulum galaxies in LEGUE are presented in Huo et al. (2013).

\subsection{Target Selections}
\label{target_selection}
The main objectives of LEGAS quasar survey are to discover more new quasars. 
Most candidates are selected based on SDSS photometry, of which the objects are required to be point-sources in SDSS DR9 (Ahn et al. 2012)  and brighter than $i = 20$. The SDSS point-spread function (PSF) magnitudes are used with Galactic extinction corrected (Schlegel et al. 1998). The bright magnitude limit, $i=16.0$, is required to avoid saturation and cross-talk in the spectra. The primary targets are selected from:
\begin{itemize}
\item {\bf optical-infrared colours} The investigations of quasar selection based on SDSS-UKIDSS/WISE colours are described in Wu \& Jia (2010) and Wu et al. (2012). These quasar candidates are selected based on their locations in multi-dimensional SDSS optical and UKIDSS/WISE infrared colour space. In LAMOST quasar survey, those selected SDSS point sources are matched with UKIDSS/Large Area Survey (LAS) DR8\footnote{The UKIDSS project is defined in Lawrence et al. (2007). UKIDSS uses the UKIRT Wide Field Camera (Casali et al. 2007) and a photometric system described in Hewett et al. (2006). The pipeline processing and science archive are described in Hambly et al. (2008).} and WISE All-Sky Data Release within a positional offset of 3\arcsec\ and 6\arcsec\, respectively. Only sources flagged as unaffected by known artifacts in all of the four bands are considered in the WISE catalog. The SDSS magnitudes referred in the following are in AB magnitudes and the UKIDSS/WISE magnitudes are in Vega magnitudes. For the SDSS-UKIDSS matched objects we select quasar candidates with the criteria of $Y-K>0.46(g-z)+0.82$ or $J-K > 0.45(i-Y-0.366)+0.64$ (Wu \& Jia 2010). For the SDSS-WISE matched objects the quasar candidates selection criteria in the regular survey are $w1-w2>0.57$ or $z-w1>0.66(g-z)+2.01$ (Wu et al. 2012). These selections mostly select quasars with redshift less than 4. 

\item{\bf data-mining algorithms} A combination of quasar selections with data-mining methods: support vector machine (SVM) classifiers (SVM; Peng et al. 2012), extreme deconvolution method (XDQSO; Bovy et al. 2011), and a kernel density estimator (KDE; Richards et al. 2009).

\end{itemize}

The primary sample described above was supplemented by the objects selected via multi-wavelength (optical-X-ray/radio) data matching, i.e., objects which have SDSS photometry in the X-ray sources of XMM-Newton, Chandra, ROSAT and radio sources in FIRST (Becker et al. 1995), NVSS (Condon et al. 1998).  Among the above selected candidates there are known quasars in SDSS, and we include some of these known quasars in LAMOST quasar survey. With the SDSS spectra of these known quasars we can test the calibrations of LAMOST spectroscopy, and investigate the spectroscopic variability of quasars.  Finally,  LAMOST quasar survey  also includes a  small number of  quasars selected by serendipitous methods.

\subsection{Spectroscopy}
\begin{figure}
\includegraphics[width=0.5\textwidth]{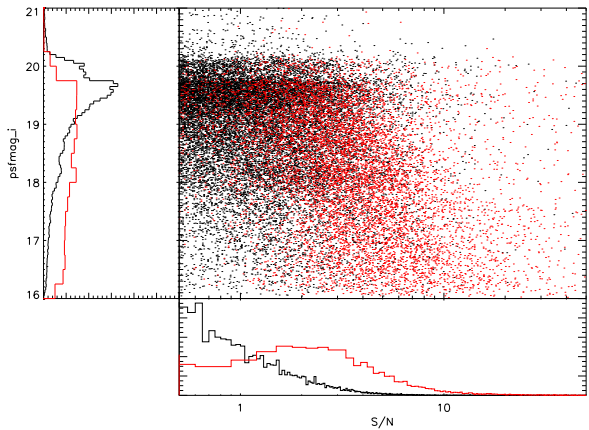}
\caption{Distributions of the S/N of LAMOST spectra  and SDSS i-band PSF magnitudes for the objects classified as UNKNOWN (black) and QSO/STAR (red) from pipeline in LAMOST LEGAS survey. \label{psfmag_all}}
\end{figure}

LAMOST LEGAS spectroscopic observations are taken in a series of at least three 30-min exposures.
There are 70,290 quasar candidates observed, with 82,625 spectra in DR1. Among them, there are $\sim$10,000 objects identified as QSO, STAR or GALAXY by  LAMOST pipeline. While the left majorities, $\sim$ 60,000, are classified as UNKNOWN. For some of these UNKNOWN the spectra are taken under non-photometric conditions, i.e., varying seeing and/or cloudy. Unstable efficiency for some fibers also contributes to the high fraction of the UNKNOWN.
The other factor is that the magnitude cutoff at $i=20$ for quasar candidates selection is too faint for LAMOST LEGAS survey. In Figure\,\ref{psfmag_all} we present the SDSS i-band PSF magnitude distributions of the observed quasar candidates. It is clear that the mean magnitude of these UNKNOWN, $i=19.09$, is around one magnitude fainter than that of the identified QSO/STAR. The median S/N per pixel in the wavelength regions of 4000-5700 and 6200-9000\,\AA\, of LAMOST spectra are also shown in the figure. Most of these UNKNOWN have spectra with relatively lower S/N, compared to QSO/STAR.

For the $\sim$10,000 identified QSO/STAR we visually inspect the spectra to secure the identifications  with Java program ASERA (Yuan et al. 2014). The misclassified objects are rejected or re-classified according to the S/N.  For quasars we only keep the spectra with at least one high S/N emission line to ensure the reliability of systematic redshift estimations. The visually inspected redshift is set by eye at the maximum of available typical quasar emission lines, i.e., \oiii, \mgii, CIII and \civ. If the redshift provided by the pipeline does not match the one estimated in ASERA, we manually correct the redshift given by the pipeline, when the spectrum has at least two emission lines available.

\begin{figure}
\includegraphics[width=0.5\textwidth]{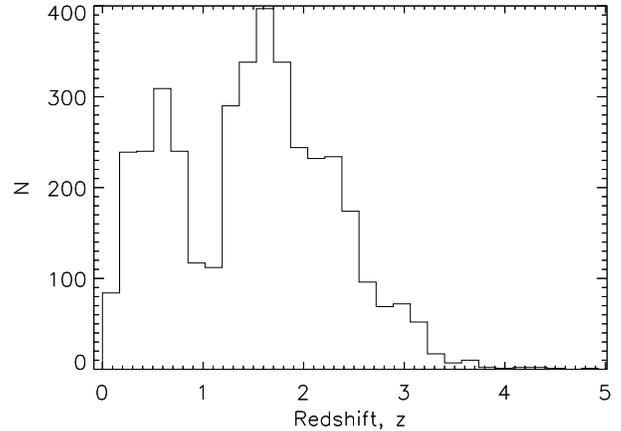}
\caption{Redshift Distributions of the reliably identified 3921 quasars in LAMOST DR1.\label{all_qso_psfmagi_z}}
\end{figure}

Finally  there are 3921 quasars  and 5355 stars with reliable identifications as secured in ASERA. 
The highest redshift of the identified quasar is 4.83 and most of the quasars are at low to median redshift with redshifts less than 3, as shown in Figure\,\ref{all_qso_psfmagi_z}.
The drop in the redshift distribution at $z \sim 1.0$  is from inefficient identifications in this redshift range as the emission line of \mgii\ moves into the overlaps of the blue and red channels. 

\begin{figure}
\includegraphics[width=0.5\textwidth]{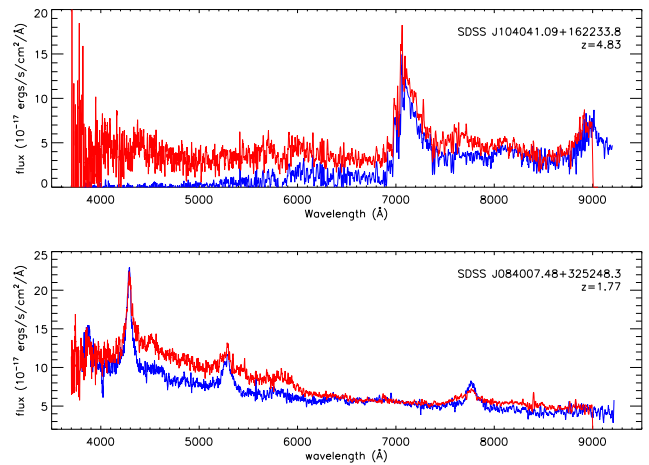}
\caption{Examples of the quasars with spectroscopic observations in LAMOST (blue) and SDSS (red). The non-calibrated LAMOST spectra are scaled to those of SDSS, and all the spectra are smoothed with a boxcar of 5 pixels for illustration.
\label{spectra_sdss_lamost}}
\end{figure}

Among the visually secured LAMOST quasars there are 2741 previously known quasars as matched with NED/SDSS DR10. For 95\% of them the redshift differences between SDSS and LAMOST are less than 0.1.  For the objects with redshift differences larger than 0.2 we visually checked the SDSS and LAMOST spectra. The differences come from  incorrect estimations of  redshifts in LAMOST  due to low S/N of the spectra or misidentifications of the emission lines. Time lags between LAMOST and SDSS spectroscopic observations range from months to decade, which are suitable to study quasar variability on both short and long timescales.  In DR1 LAMOST release the spectra with non-flux-calibrated. Thus it is much more applicable to study line profile variabilities of quasars, instead of flux variations.
In Figure\,\ref{spectra_sdss_lamost} we present LAMOST and SDSS spectra as example of quasars with consistent redshifts, including the one with highest redshift ($z=4.83$) identified by LAMOST. 

\begin{figure}
\includegraphics[width=0.5\textwidth]{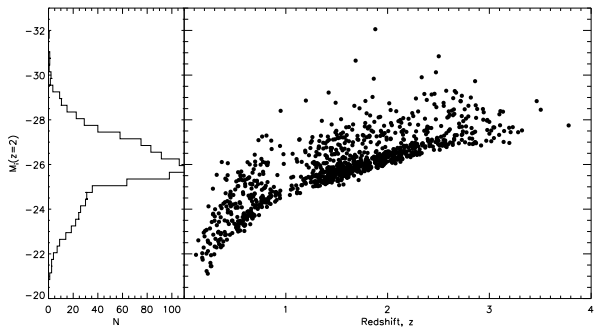}
\caption{Distributions of the new quasars discovered in LAMOST DR1 in luminosity-redshift space, where the left panel shows the luminosity histograms.
\label{new_quasar_Mi_Z}}
\end{figure}

The main purpose of this paper is to compile a catalog for the 1180 new quasars discovered in LAMOST DR1. The redshifts and i-band absolute magnitudes for these quasars are shown in Figure\,\ref{new_quasar_Mi_Z}. Here luminosity is indicated using the (continuum and emission line) K-corrected, i-band absolute magnitude, $M_{i}(z=2)$, normalized at $z=2$ (Richard et al. 2006).
In the following we present detailed descriptions about the emission line measurements and black hole mass estimations for these quasars.

\section{Spectral Measurements}

We measure the properties of  \ha, \hb, \mgii, and \civ\ lines with spectral fittings. These lines are the typical, strong emission lines of quasars and calibrated as virial black hole mass estimators. The fits are based on the MPFIT package(Markwardt 2009), which performs $\chi^{2}$-minimization using the Levenberg$\sbond$Marquardt technique.
Details of the spectral fitting in the optical and near-ultraviolet regions have been described in the literatures (e.g., Dong et al. 2008, Wang et al. 2009a, Ho et al. 2012).  The cross-over between the red and blue channels, wavelength range of 5700-6200 \AA\ in observed frame, is masked out. All the LAMOST spectra are corrected for Galactic extinction using the extinction map of Schlegel et al. (1998)  and the reddening curve of Fitzpatrick (1999).  The spectra are shifted back to their rest frames in the fitting.

For each line we first fit a local pseudo-continuum, i.e., a simple power-law or a power-law plus an \feii\ multiplet emissions, to the wavelength regions least contaminated by the broad line emissions. The pseudo-continuum fitting windows for each line are referred to those used in Shen et al. (2011). The parameters of normalization $\alpha$ and slope $\beta$ of the power-law continuum are set to be free. The parameters of the \feii\  multiplet emissions, i.e., normalization, velocity shift  relative to the systemic redshift, and broadening velocity, are also set to be free. The fitted pseudo-continuum emissions are subtracted from the spectra and the leftover emission-line spectra are fitted with Gaussian and/or Gaussian-Hermite profiles. Below we describe the detailed fitting procedures for each line.

\subsection{\ha\ and \hb }
We fit the spectra of quasars with $z\la0.30$ for \ha, and $z\la0.75$ for \hb. The fitting windows of power-law plus \feii\ emissions are [6000, 6250] \AA\ and [6800, 7000] \AA\ for \ha, [4435, 4700] \AA\ and [5100, 5535] \AA\ for \hb. The template of optical \feii\ emission is from \citet[][]{veron04}. Gaussian function is used to model the emission lines. Both \ha\ and \hb\ emission lines are modelled with one narrow  and one broad component. The broad component is fitted with as many Gaussians as statistically justified (see Dong et al. 2008 for details). All narrow emission lines, except the \oiii\ doublet, are fitted with a single Gaussian. The upper limits of the full width at half-maximum (FWHM) for the narrow lines  are set to be 1200 \kmps.

\begin{figure}
\includegraphics[width=0.5\textwidth]{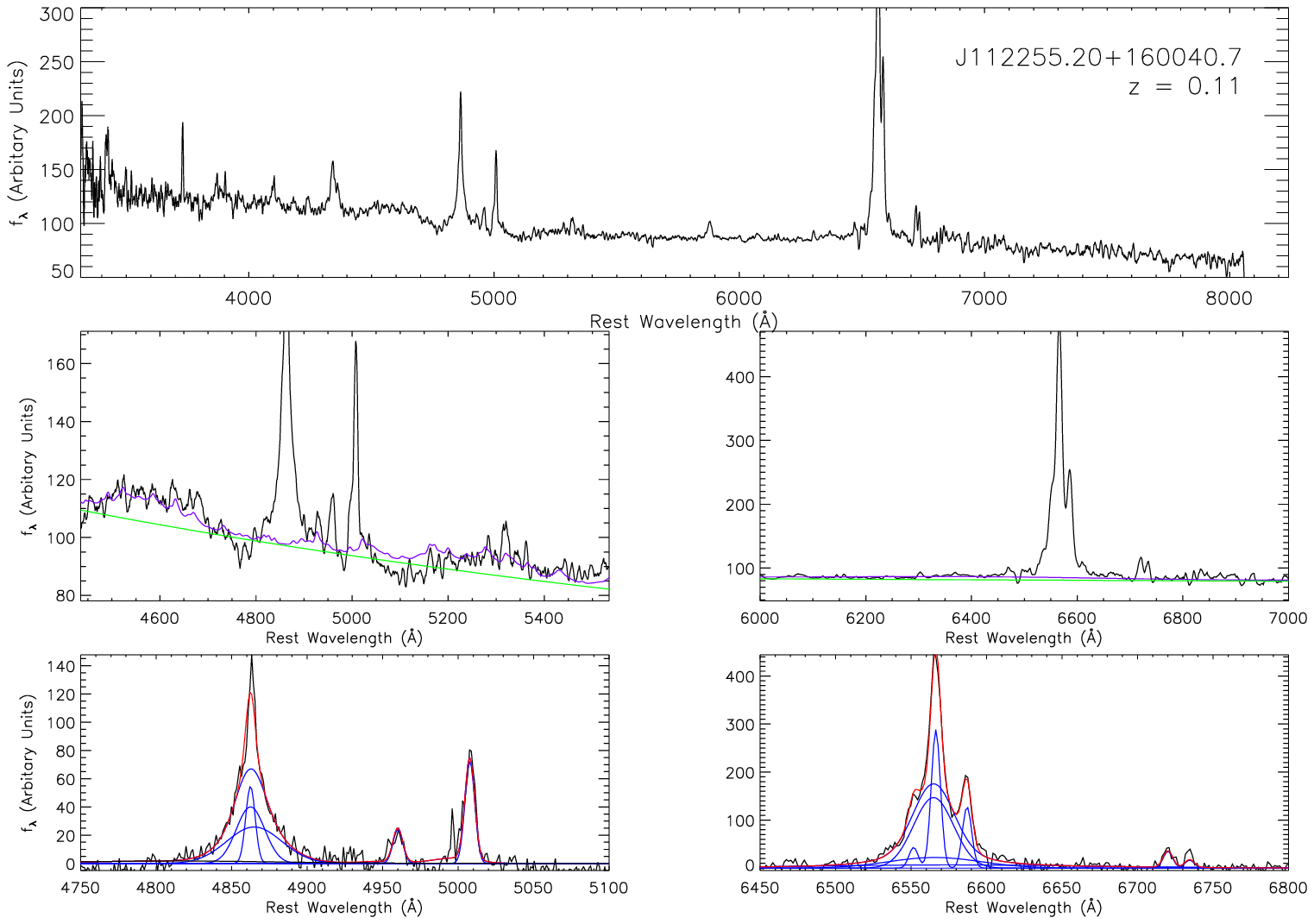}
\caption{Representative example of the fitting applied to \hb\ + \oiii\ and \ha\ + \nii\ ranges. Upper panel is the observed spectrum; middle panels show the pseudo-continuum  fitting with power-law (green) and \feii\ multiplets (cyan lines indicate the \feii\ plus power-law emission); bottom panels are the results of deblending 
procedures applied to \hb\ + \oiii\ and \ha\ + \nii\ emission lines, in which combination of models (red) and individual components are shown (blue). 
The observed spectra are smoothed with a boxcar of 5 pixels for the illustration.
\label{example_hb}}
\end{figure}

Each line of the \oiii\ $\lambda\lambda$4959, 5007 doublet is modelled with two Gaussians, one accounting for the line core and the other for a possible blue wing as seen in many objects (e.g., Greene \& Ho 2005a, Komossa et al. 2008, Wang et al. 2009b). The doublet  are assumed to have the same redshifts and profiles, with flux ratio fixed to the theoretical value of 3. Velocity offsets and line widths of the doublet core component are tied to those of narrow \hb\ component.  The velocity offsets  and line widths of \nii\ $\lambda\lambda$6548, 6584  and \sii\ $\lambda\lambda$6717, 6731 are tied to those of  \ha\ narrow component. The relative flux ratio of the two \nii\ components is fixed to 2.96. We present the fittings to \hb\ and \ha\ lines in Figure\,\ref{example_hb} as illustrations.

\subsection{\mgii\ and \civ}
We fit the spectra of quasars with $0.40\la z \la 1.76$ for \mgii\ line, and $z \ga 1.58$ for \civ.  The pseudo-continuum fitting windows are [2200, 2700] \AA\ and [2900, 3090] \AA\ for \mgii, [1445, 1465] \AA\ and [1700, 1705] \AA\ for \civ. Ultraviolet \feii\ template from \citet[][]{tsuzuki06} is included in the continuum modelling of \mgii, while not for \civ.  This mainly out of  the consideration that  \feii\ emissions are usually not strong for \civ, and  inclusion of \feii\ emission will add uncertainties to the fitted parameters based on data quality of LAMOST.

\begin{figure}
\includegraphics[width=0.5\textwidth]{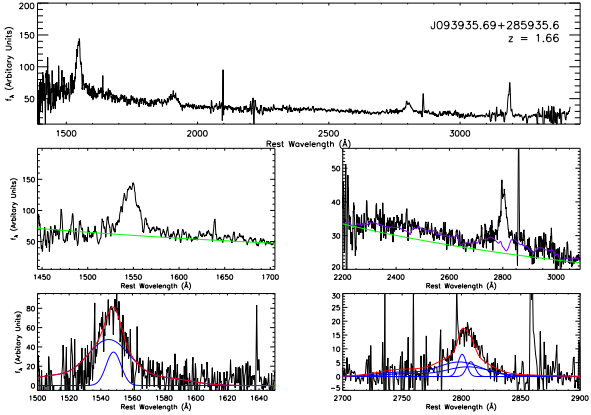}
\caption{ Representative example of the fitting applied to \mgii\ and \civ\ lines.  Upper panel is the observed spectrum; middle panels show the pseudo-continuum  fitting with power-law (green) and \feii\ multiplets (cyan lines indicate the \feii\ plus power-law emission); bottom panels are the results of deblending 
procedures applied to \mgii\ and \civ\ emission lines, in which combination of models (red) and individual components are shown (blue). 
The observed spectra are smoothed with a boxcar of 5 pixels for the illustration.
\label{example_MgII}}
\end{figure}

Each of the \mgii\ $\lambda\lambda$2796, 2803 doublet are modelled with one narrow and broad component. The narrow component is modelled with a single Gaussian, and the upper limit of FWHM is set to be 900 \kmps (Wang et al. 2009a).  We model the broad component with a truncated five-parameter Gauss-Hermite series (van der Marel\&Franx 1993; see also Salviander et al. 2007). The broad components of \mgii\ doublet lines are set to have the same profile, with flux ratio set to be between 2:1 and 1:1 (Laor et al. 1997). The same prescription is applied to the narrow components.  For \civ\  we model the line with one Gaussian and one Gaussian-Hermit function to obtain a smooth realization of the line profile. We do not  set upper limit for FWHM of the Gaussian function since there are still debates whether a strong narrow component exists for \civ\ line (Baskin \& Laor  2005). We present the fittings to \mgii\ and \civ\ lines in Figure\,\ref{example_MgII} as illustrations.

The fittings to  \mgii\ and \civ\ emission lines are frequently affected by narrow or broad absorption features. The effects on the fittings from broad absorption trough are hard to be reduced in automatic fitting procedure, while the impact of narrow or moderate absorption features can be alleviated (Shen et al. 2011). We mask out 3$\sigma$ outliers below the 20 pixel boxcar-smoothed spectrum to reduce the effects of narrow or moderate absorptions on the \civ\ and \mgii\ emission line fittings. 

\subsection{Reliability of Spectral Fittings}
We visually inspect the fitting results for each object. Most of the fits to spectrum with high S/N are acceptable, as shown in Figure\,\ref{example_hb} and Figure\,\ref{example_MgII}. The bad fits include spectra with low S/N, too few good pixels in the fitting region, or objects with peculiar continuum and emission line properties, such as broad-absorption line quasars and disk emitters.
A flag is given for each line based on visual inspections to indicate the quality of spectral fitting.  flag = 0 means that the fitting is acceptable and the inferred parameters are mostly reliable, while flag = -1 means that the line measurements maybe spurious.

\section{Continuum Luminosity and Virial Black Hole Mass}

We fit the five-bands SDSS photometry with model spectra  to estimate continuum luminosity for the new quasars.  
At first step we extract SDSS PSF magnitudes only for the quasars which have clean photometry\footnote{http://www.sdss3.org/dr10/tutorials/flags.php\#clean}  in SDSS database. The Galactic extinction corrected SDSS $ugriz$ magnitudes are converted to the corresponding magnitudes on the AB system\footnote{http://www.sdss3.org/dr9/algorithms/fluxcal.php\#SDSStoAB}. These AB magnitudes are then turned into the flux density, F$_{\lambda}$, at the effective wavelength of each filter. 
We then fit the five data points, F$_{\lambda}$, of each quasar at rest frame with model spectra composed of two components:  continuum emissions and  line emissions.  
The features of this two components are referred to those of the composite quasar spectra in Vanden Berk (2001) (hereafter VB01).  As described in VB01 the continuum of composite quasar spectra is 
well fitted by broken power laws, and  the features left after the subtraction of the continuum in the median composite quasar spectra are taken as the template of line emissions in our model.  

In our fitting the indices of the broken power law and normalization of line emissions are set to be free parameters.
The break of the two power laws is set to the value of 4661\,\AA,  which is same to the value inferred from median composite spectra in VB01.
Normalization of the continuum is tied to the mean value of the total emissions in the regions of 1350-1365 and 4200-4230\,\AA. The values of these two regions are taken from VB01, in which they fit the continuum.
At blueward of Ly$\alpha$ we only scale the composite median quasar spectra. The fitting is performed by minimizing the chi-squared, $\chi^{2}$.

\begin{figure}
\includegraphics[width=0.5\textwidth]{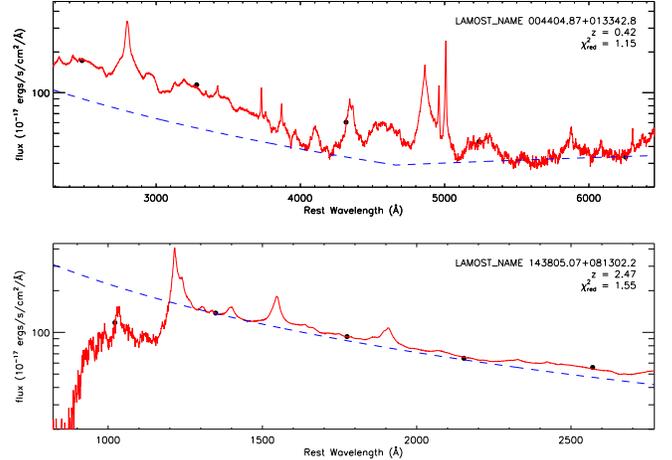}
\caption{Examples of estimation of continuum luminosity with model fitting to SDSS photometry (black points). The fitted total model spectra (red lines) and power law continuum (blue dashed lines) are shown.
\label{example_model_sdss}}
\end{figure}

We present examples of the modelling in Figure\,\ref{example_model_sdss}. As shown in the figure, the continuum model of quasars with $z\ga1$ is actually a single power law since for these quasars the value of break of the two power laws, set in our model, moves out of the LAMOST wavelength region.  For these quasars the degrees of freedom (dof) is 2, which is one more compared to that of the quasars with $z<1$. The mean frequency power-law index inferred from the fitting, $<\alpha_{\nu}>$, is -- 0.43 from $\sim$ 1300 to 4661\,\AA\ and --1.60 redward of $\sim$ 4661\,\AA, which are in good agreement with the values found in composite quasar spectra in VB01.

\begin{figure}
\includegraphics[width=0.5\textwidth]{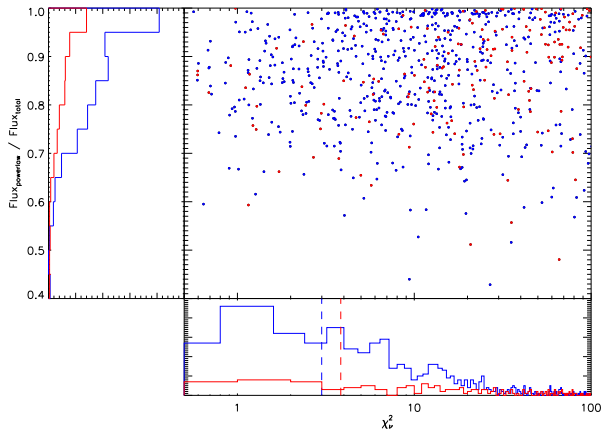}
\caption{Distribution of reduced chi-squared, $\chi^{2}_{\nu}$, and flux ratio of continuum to total emission in the data fitting range. Red points are the quasars with $z < 1$, and blue points are the rests. The vertical dashed lines label the values of 3.841 and 2.996 of $\chi^{2}_{\nu, 0.05}$, which are  corresponding to the probability of 0.05 for obtaining a value $\chi^{2}_{\nu} > \chi^{2}_{\nu, 0.05}$ with dof of 1 and 2. \label{redchi_ratio}}
\end{figure}

The distribution of reduced chi-squared, $\chi^{2}_{\nu}$, is shown in Figure\,\ref{redchi_ratio}. In the figure we mark the values of reduced chi-squared, $\chi^{2}_{\nu, 0.05}$, which are  corresponding to the probability of 0.05 for obtaining a value $\chi^{2}_{\nu} > \chi^{2}_{\nu, 0.05}$. The  $\chi^{2}_{\nu, 0.05}$ is 3.841 with dof of 1 and 2.996 with dof of 2.  The fittings with $\chi^{2}_{\nu} < \chi^{2}_{\nu, 0.05}$  are good fits in statistical significance level. However, not all the fits with relatively larger $\chi^{2}_{\nu}$ are unacceptable based on visual inspections. In Figure\,\ref{redchi_ratio} we also present the distribution of the fraction of continuum emission accumulated in the data fitting wavelength range (only from rest-frame wavelength of 1200\,\AA\ if data extends to blueward of Ly$\alpha$). 
The fractions of continuum emission are mostly in the range of greater than 60\%, dispersed around the values (84\%--88\%) of those in VB01.
The fractions with extreme high values are mostly derived from the quasars with bad fitting. In a loose constraint we reject the fittings with fraction values greater than 0.989. 

With inferred continuum luminosity, i.e., $L_{5100}$, $L_{1350}$ and, $L_{3000}$, and corresponding broad line width we can estimate the virial black hole masses with 
\begin{align} % requires amsmath; align* for no eq. number
   {\rm log}\left(\frac{M_{BH}}{M_{\sun}}\right) = \alpha + \beta\, {\rm log}\left(\frac{\lambda L_{\lambda}}{10^{44} \,{\rm erg\, s^{-1}}}\right) + \gamma\, {\rm log}\left(\frac{\rm FWHM}{\rm km\, s^{-1}}\right)
\end{align}
The coefficients $\alpha$, $\beta$, and  $\gamma$ are empirically calibrated and the values differ in the calibrations with different measurements of emission lines and samples (e.g., McLure \& Dunlop 2004; Greene \& Ho 2005b; Vestergaard \& Peterson 2006; Wang et al. 2009a; Zuo et al. 2014). In this paper for \hb-based and \civ-based estimates we use the virial black hole mass calibrations  from Vestergaard \& Peterson (2006), and for \mgii-based from Wang et al.(2009a). The FWHM of broad Mg II $\lambda2796$, and FWHM of whole \civ\ emission are used in the estimations following the FWHM definitions adopted in these calibrations. The black hole masses are estimated only for those quasars with inferred continuum luminosity and broad line widths both flagged as reliable.

\begin{figure}
\includegraphics[width=0.5\textwidth]{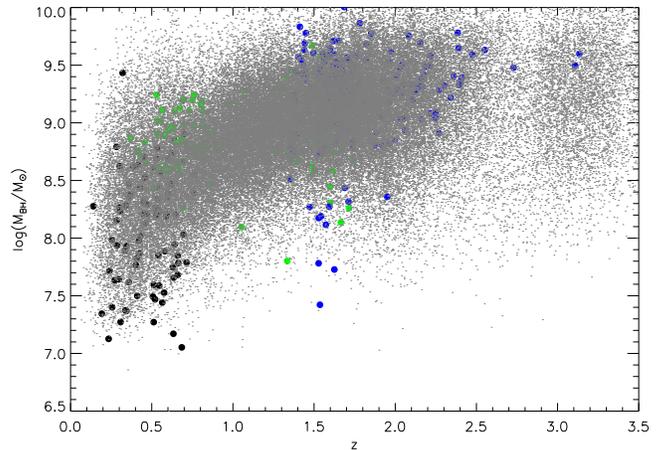}
\caption{Filled circles are the new quasars discovered in LAMOST DR1 with estimated black hole masses based on emission lines of \hb\ (black), Mg II (green), and C IV (blue) .
Small grey dots are SDSS quasars from Shen et al.(2011). \label{bh_mass}}
\end{figure}

We notify that the continuum luminosity estimations described above introduce additional uncertainties to the black hole mass estimations compared to the ones from single-epoch spectra. The uncertainties come from model fitting to the SDSS photometric data and intrinsic variations of quasar luminosity. Time lags between SDSS photometric data and LAMOST spectra are $\sim$ 10 years. On this timescale the optical/ultraviolet emissions of quasars are variable with amplitudes of 0.1-0.2 mag (e.g., Vander Berk et al. 2004; Ai et al. 2013; Zuo et al. 2012).  
%It is hard to estimate the introduced uncertainties quantitatively. 
To justify our estimations we present the apparent trend of the black hole masses with redshifts in Figure\,\ref{bh_mass}. It is evident that the distributions of the black hole masses for these LAMSOT quasars are overlapped with those of SDSS DR7 quasars from Shen et al. (2011). The result supports that our continuum modelling is mostly valid and the estimated black hole masses can be taken as proxy for these new quasars. 

\section{The Spectral Catalog}

We present a compiled catalog for the new quasars identified in LAMOST DR1.  All measured quantities are tabulated in the online catalog  as binary FITS table file at LAMOST public website. We cross-correlate our quasar catalog with objects in FIRST radio survey (Becket et al. 1995) within 2.0$\arcsec$ radius  and list the FIRST peak flux density.  A summary of the information contained in each of the columns is described in Table\,\ref{table_format}. Below we describe the specifics of the cataloged quantities.\\
1. LAMOST object designation: LAMOST Jhhmmss.ss+ddmmss.s(J2000, truncated coordinates). The `LAMOST J' for each entry is dropped. \\
2-3. Right Ascension and Declination (in decimal degrees, J2000).\\
4-7. Information of spectroscopic observation:  Modified Julian date (MJD), spectroscopic plan identification (planID), spectrograph identification (spID), and spectroscopic fiber number (fiberID)). These four numbers are unique for each spectrum named in the format of $spec-MJD-planID\_spID-fiberID.fits$. Only information of the spectrum with highest S/N is given if more than one spectroscopic observations available for the quasar. \\
8. Redshift. \\
9. ZWARNING flag  based on visual inspections. 1 = not robust.\\
10. $M_{i}(z=2)$: absolute i-band magnitude. K-corrected to $z = 2$ following Richards et al. (2006).\\
11. NSPECOBS: number of spectroscopic observations for the quasar.\\
12. SNR\_SPEC: median S/N per pixel in the wavelength regions of 4000-5700 and 6200-9000\,\AA. \\
13-19. FWHM, rest-frame equivalent width (EW) for broad \ha, narrow \ha, [NII]6584, [SII]6717, and [SII]6731.\\
20. Rest-frame equivalent width of iron emissions in 6000-6500 \AA. \\
21-22. Number of good pixels and median S/N per pixel for the \ha\ region in 6400-6765 \AA. \\
23. Flag indicates reliability of the emission line fitting results in \ha\ region upon visual inspections. -1 = unacceptable; 0 = acceptable. \\
24-29. FWHM, rest-frame equivalent width for broad \hb, narrow\hb, [OIII]4959, [OIII]5007, fitted in the 4750-4950 \AA\ region. \\
30. Rest-frame equivalent width of iron lines in 4435-4685 \AA. \\
31-32. Number of good pixels and median S/N per pixel for the \hb\ region in 4750-4950 \AA. \\
33. Flag indicate reliability of the emission line fitting results in \hb\ upon visual inspections.  -1 = unacceptable; 0 = acceptable. \\
34-35. FWHM, rest-frame equivalent width for total MgII emission line. \\
36-37. FWHM, rest-frame equivalent width for total broad MgII lines. \\
38-39. FWHM for broad and narrow MgII 2796 emission line. \\
40. Rest-frame equivalent width of iron lines in 2200-3090 \AA . \\
41-42. Number of good pixels and median S/N per pixel for the MgII region in 2700-2900\,\AA. \\
43. Flag indicates reliability of the emission line fitting results in MgII upon visual inspections. -1 = unacceptable; 0 = acceptable. \\
44-45. FWHM, rest-frame equivalent width for total CIV emission line. \\
46-47. FWHM, rest-frame equivalent width for Gaussian-Hermit component of CIV line. \\
48-49. FWHM, rest-frame equivalent width for Gaussian component of CIV line. \\
50-51. Number of good pixels and median S/N per pixel for the CIV region in 1500-1600\,\AA. \\
52. Flag indicates reliability of the emission line fitting results in CIV upon visual inspections. -1 = unacceptable; 0 = acceptable. \\
53-54. Wavelength power-law index, $\alpha_\lambda$, from $\sim$ 1300 to 4661\,\AA, and redward of 4661\,\AA. \\
55. Reduced chi-square in SDSS photometry modelling; -- 99 if not fitted. \\
56-58. Continuum luminosity at 5100\,\AA, 3000\,\AA, and 1350\,\AA.\\
59-61. Virial black hole masses with calibrations of \hb\ (VP06), \mgii\ (Wang08), and \civ\ (VP06).\\
62. FIRST peak flux density at 20 cm. \\
62. Flag indicates whether a broad absorption trough present in the spectra by visual inspection. 1 is set if visually identified as broad absorption quasar. \\
64. Flag set as SVM, XD, or Richards09; mark quasars selected with corresponding data-mining method. \\
65. Flag set as W, or U; mark quasars selected with SDSS/UKIDSS, or SDSS/WISE colours. \\
66. Flag set as Xray, RADIO, or S; mark quasars selected by X-ray, radio, or other serendipitous algorithms. \\

\section{Conclusion and Discussion}
We present our preliminary results of LEGAS quasar survey in LAMOST DR1, which includes products from pilot survey and the first year regular survey. 
We compile emission line measurements around the \ha, \hb, \mgii, and \civ\ regions for the 1180 new quasars discovered in the survey.  We also compile the virial black hole mass estimates, and flags indicating the selection methods, broad absorption line quasars. The catalog and spectra for these quasars are available online.

\begin{figure}
\includegraphics[width=0.5\textwidth]{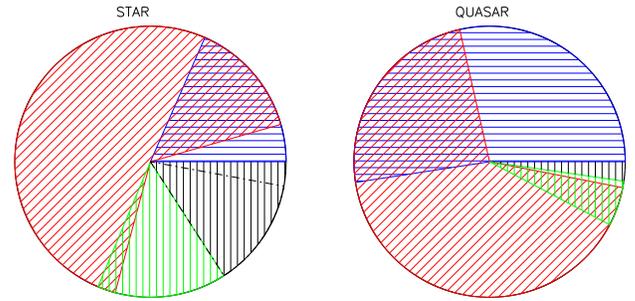}
\caption{Illustration of the fractions of objects selected from optical-infrared colours (blue), data-mining algorithms (red), multi-wavelengths (optical-X-ray/radio) matching (green), and serendipitous algorithms (black) for the 5355 stars and 3921 quasars in LAMOST DR1. The stars in the area of dash-dotted lines are observed in pilot survey, of which the selections are a bit different with those in regular survey. \label{percent}}
\end{figure}

\begin{figure}
\includegraphics[width=0.5\textwidth]{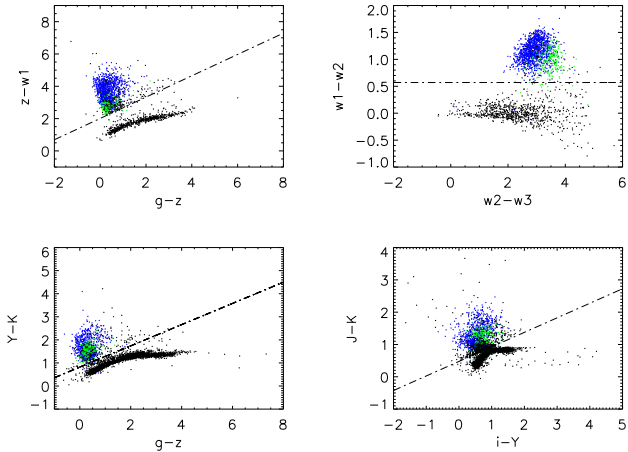}
\caption{Distributions of 3921 quasars and 5355 stars (black dots) in SDSS-WISE/UKIDSS colour diagram. Blue dots are the quasars with $z<2.2$, and green are $z\ga2.2$. The dashed dotted lines indicate criteria used in regular survey.
\label{all_quasar_color}}
\end{figure}

The candidates in LEGAS quasar survey are selected from optical-infrared colors, data-mining algorithms, and multi-wavelength (optical--X-ray/radio) data matching. 
 Object fractions selected from these methods for the reliably identified 5355 stars and 3921 quasars are shown in Figure\,\ref{percent}.  
 %There are 96\% of the quasars selected from optical-infrared colours and data-mining algorithms. 
 There are 28\% of the quasars selected with optical-infrared colours independently, indicating that the method is quite promising for completeness of low redshift quasars and searching for high redshift quasars (Wu et al. 2012, Wu et al. 2015).  We have improved candidates selections in regular survey compared to those in pilot survey, in which relatively larger fraction of stars are selected, as shown in the figure. We present the SDSS-WISE/UKIDSS colour-colour distributions of these quasars/stars in Figure\,\ref{all_quasar_color}, in which the criteria used in regular survey are shown in dash-dotted lines.
 
In DR1, a significant high fraction ($\sim$ 80\%) of the observed candidates can not be identified by the pipeline. Most of these UNKNOWN have spectra with low S/N, due to non-photometric conditions, unstable fiber efficiency, and fainter magnitude cutoff at $i = 20$ for quasar candidate selections. To increase quasar identification efficiency we have adjusted the magnitude cutoff to $i=19$ in the on-going survey.

We provide the information of LAMOST spectroscopic observations and selection methods for the 2741 known quasars in Table\,\ref{known_quasar}. 
With LAMOST and SDSS spectra of these quasars, we can investigate quasar spectral variability on timescales from months to decade.
It is worthwhile to launch follow-up studies for the interesting quasars discovered in DR1, such as those with broad absorption troughs in \civ\ and/or \mgii,  double-peaked Balmer line profiles, etc.  LAMOST DR1 and the on-going survey will provide valuable data in the studies of quasars.

%%%%%%%%%%%%%%%%%%%%%%%%%%%%%%%%%%%%%%%%%%%%%%%%%%%%%%%%%%%%
\acknowledgments

This work is supported by NSFC grants 11103071, 11373008, 11033007, 2014CB845705, and State Key Development Program for Basic Research of China (No. 2013CB834900 and 2015CB857000).
Guoshoujing Telescope (the Large Sky Area Multi-Object Fiber Spectroscopic Telescope LAMOST) is a National Major Scientific Project built by the Chinese Academy of Sciences.Funding for the project has been provided by the National Development and Reform Commission.LAMOST is operated and managed by the National Astronomical Observatories, Chinese Academy of Sciences. 

This publication makes use of data products from the Wide-field
Infrared Survey Explorer, which is a joint project of the
University of California, Los Angeles, and the Jet Propulsion
Laboratory/California Institute of Technology, funded by the National Aeronautics and Space Administration.

Funding for the Sloan Digital Sky Survey IV has been provided by the Alfred P. Sloan Foundation and the Participating Institutions. SDSS-IV acknowledges support and resources from the Center for High-Performance Computing at the University of Utah. The SDSS web site is www.sdss.org.

SDSS-IV is managed by the Astrophysical Research Consortium for the Participating Institutions of the SDSS Collaboration including the Carnegie Institution for Science, Carnegie Mellon University, the Chilean Participation Group, Harvard-Smithsonian Center for Astrophysics, Instituto de Astrof¨ªsica de Canarias, The Johns Hopkins University, Kavli Institute for the Physics and Mathematics of the Universe (IPMU) / University of Tokyo, Lawrence Berkeley National Laboratory, Leibniz Institut f¨¹r Astrophysik Potsdam (AIP), Max-Planck-Institut f¨¹r Astrophysik (MPA Garching), Max-Planck-Institut f¨¹r Extraterrestrische Physik (MPE), Max-Planck-Institut f¨¹r Astronomie (MPIA Heidelberg), National Astronomical Observatory of China, New Mexico State University, New York University, The Ohio State University, Pennsylvania State University, Shanghai Astronomical Observatory, United Kingdom Participation Group, Universidad Nacional Aut¨®noma de M¨¦xico, University of Arizona, University of Colorado Boulder, University of Portsmouth, University of Utah, University of Washington, University of Wisconsin, Vanderbilt University, and Yale University.

%%%%%%%%%%%%%%%%%%%%%%%%%%%%%%%%%%%%%%%%%%%%%%%%%%%%%%%%%%%%%%%%%%%%%%%%%%%%%%%%%%%%%%%%%%%%%%%%%%%%
%%%%%%%%%%%%%%%%%%%%%%%%%%%%%%%%%%%%%%%%%%%%%%%%%%%%%%%%%%%%%%%%%%%%%%%%%%%%%%%%%%%%%%%%%%%%%%%%%%%%

%%%%%%%%%%%%%%%%%%%%%%%%%%%%%%%%%%%%%%%%%%%%%%%%%%%%%%%%%%%%%%%%%%%%%%%%%%%%%%%%%%%%%%%%%%%%%%%%%%
%%%%%%%%%%%%%%%%%%%%%%%%%%%%%%%%%%%%%%%%%%%%%%%%%%%%%%%%%%%%%%%%%%%%%%%%%%%%%%%%%%%%%%%%%%%%%%%%%

\clearpage
%\begin{center}
%\tiny
%LongTables
\begin{deluxetable*}{clcc}
\tabletypesize{\scriptsize} 
\tablecaption{Catalog format for the 1180 new quasars discovered in LAMOST DR1\label{table_format}}
\tablehead{\colhead{Column} &  \colhead{Name} &  \colhead{Format} &  \colhead{Description}}  
\startdata
1   & LAMOST\_NAME  &  STRING  &  LAMOST DR1 designation hhmmss.ss+ddmmss (J2000) \\
2   & RA            &  DOUBLE  &  Right ascension in decimal degrees (J2000) \\
3   & DEC           &  DOUBLE  &  Declination in decimal degrees (J2000) \\
4   & MJD           &  LONG    &  MJD of spectroscopic observation \\
5   & PLANID       &  STRING  &  Spectroscopic plan identification \\
6   & SPID          &  STRING  &  Spectrograph identification \\
7   & FIBERID       &  LONG    &  Spectroscopic fiber number \\
\hline
8   & Z\_VI          &  DOUBLE  &  Redshift from visual inspection \\
9   & ZWARNING        &  LONG    &  ZWARNING flag from visual inspection \\
%10  & Z\_PIPE        &  DOUBLE  &  Redshift from LAMOST pipeline \\
%11  & Z\_PIPE\_ERR   &  DOUBLE  &  Errors on LAMOST pipeline redshift \\
10  & MI\_Z2         &  DOUBLE  &  M$_{i} (z=2)$, K-corrected to $z=2$ following Richards et al. (2006)] \\
11  & NSPECOBS	              &  LONG	&  Number of spectroscopic observations \\
12  & SNR\_SPEC               &  DOUBLE &  Median S/N per pixel of the spectrum \\
\hline
13  & FWHM\_BROAD\_HA	      &  DOUBLE	&  FWHM of broad \ha\ in \kmps \\
%16  & FWHM\_BROAD\_HA\_ERR    &  DOUBLE	&  Uncertainty on the broad \ha\ in \kmps \\
14	& EW\_BROAD\_HA	          &  DOUBLE	&  Rest frame equivalent width of broad \ha\ in \AA  \\
%17	& EW\_BROAD\_HA\_ERR	  &  DOUBLE	&  Uncertainty in EW\_BROAD\_HA in \AA  \\
15	& FWHM\_NARROW\_HA	      &  DOUBLE	&  FWHM of narrow \ha\ in \kmps \\
%20	& FWHM\_NARROW\_HA\_ERR	  &  DOUBLE	&  Uncertainty on the narrow \ha\ in \kmps \\
16  & EW\_NARROW\_HA	      &  DOUBLE	&  Rest frame equivalent width of narrow \ha\ in \AA \\
%22	& EW\_NARROW\_HA\_ERR	  &  DOUBLE	&  Uncertainty in EW\_BROAD\_HA in \AA \\
17	& EW\_NII\_6585           &  DOUBLE	&  Rest frame equivalent width of [NII]6584 in \AA \\
%24	& EW\_NII\_6585\_ERR	  &  DOUBLE	&  Uncertainty in EW\_NII\_6585 in \AA \\
18	& EW\_SII\_6718   	      &  DOUBLE	&  Rest frame equivalent width of [SII]6717 in \AA \\
%26	& EW\_SII\_6718\_ERR	  &  DOUBLE	&  Uncertainty in EW\_SII\_6718 in \AA \\
19	& EW\_SII\_6732   	      &  DOUBLE	&  Rest frame equivalent width of [SII]6731 in \AA  \\
%28	& EW\_SII\_6732\_ERR	  &  DOUBLE	&  Uncertainty in EW\_SII\_6732 in \AA \\
20	& EW\_FE\_HA	          &  DOUBLE	&  Rest frame equivalent width of Fe within 6000-6500\AA\ in \AA \\
21	& LINE\_NPIX\_HA	      &  LONG	&  Number of good pixels for the rest frame 6400-6765 \AA  \\
22	& LINE\_MED\_SN\_HA	      &  DOUBLE	&  Median S/N per pixel for the restframe 6400-6765 \AA   \\
%25	& LINE\_REDCHI2\_HA	      &  DOUBLE	&  Reduced $\chi^{2}$ for the \ha\ fit; -1 if not fitted  \\
23  & LINE\_FLAG\_HA          &  LONG   &  Flag for the quality in \ha\ fitting \\
\hline
24  & FWHM\_BROAD\_HB	      &  DOUBLE	&  FWHM of broad \hb\ in \kmps \\
%35  & FWHM\_BROAD\_HB\_ERR    &  DOUBLE	&  Uncertainty on the broad \hb\ in \kmps \\
25	& EW\_BROAD\_HB	          &  DOUBLE	&  Rest frame equivalent width of broad \hb\ in \AA  \\
%37	& EW\_BROAD\_HB\_ERR	  &  DOUBLE	&  Uncertainty in EW\_BROAD\_HB in \AA  \\
26	& FWHM\_NARROW\_HB	      &  DOUBLE	&  FWHM of narrow \hb\ in \kmps \\
%39	& FWHM\_NARROW\_HB\_ERR	  &  DOUBLE	&  Uncertainty on the narrow \hb\ in \kmps \\
27  & EW\_NARROW\_HB	      &  DOUBLE	&  Rest frame equivalent width of narrow \hb\ in \AA \\
%41	& EW\_NARROW\_HB\_ERR	  &  DOUBLE	&  Uncertainty in EW\_BROAD\_HB in \AA \\
28	& EW\_OIII\_4959          &  DOUBLE	&  Rest frame equivalent width of [OIII]4959 in \AA  \\
%43	& EW\_OIII\_4959\_ERR	  &  DOUBLE	&  Uncertainty in EW\_OIII\_4959 in \AA \\
29	& EW\_OIII\_5007	      &  DOUBLE	&  Rest frame equivalent width of [OIII]5007 in \AA \\
%45	& EW\_OIII\_5007\_ERR     &	 DOUBLE	&  Uncertainty in EW\_OIII\_5007 in \AA \\
30	& EW\_FE\_HB              &  DOUBLE	&  Rest frame equivalent width of Fe within 4435-4685A in \AA \\
31	& LINE\_NPIX\_HB	      &  LONG	&  Number of good pixels for the rest frame 4750-4950 \AA  \\
32	& LINE\_MED\_SN\_HB	      &  DOUBLE	&  Median S/N per pixel for the res tframe 4750-4950 \AA  \\
%35	& LINE\_REDCHI2\_HB	      &  DOUBLE	&  Reduced $\chi^{2}$ for the \hb\ fit; -1 if not fitted \\
33  & LINE\_FLAG\_HB          &  LONG   &  Flag for the quality in \hb\ fitting\\
\hline
34	& FWHM\_MGII              &  DOUBLE	&  FWHM of the whole MgII emission line in \kmps \\
%52	& FWHM\_MGII\_ERR	      &  DOUBLE	&  Uncertainty in FWHM\_MgII in \kmps \\
35	& EW\_MGII                &	 DOUBLE	&  Rest frame equivalent width of the whole MgII \AA \\
%54	& EW\_MGII\_ERR	          &  DOUBLE	&  Uncertainty in EW\_MgII in \AA \\
36	& FWHM\_BROAD\_MGII	      &  DOUBLE	&  FWHM of the whole broad MgII in \kmps \\
%56	& FWHM\_BROAD\_MGII\_ERR  &  DOUBLE	&  Uncertainty in FWHM\_broad\_MgII in \kmps  \\
37	& EW\_BROAD\_MGII	      &  DOUBLE	&  Rest frame equivalent width of the whole broad MgII in \AA \\
%58	& EW\_BROAD\_MGII\_ERR	  &  DOUBLE	&  Uncertainty in EW\_broad\_MgII in \AA \\
38  & FWHM\_BROAD\_MGII\_2796 &  DOUBLE &  FWHM of the broad MgII 2796 in \kmps \\
39  & FWHM\_NARROW\_MGII\_2796      &  DOUBLE &  FWHM of the narrow MgII 2796 in \kmps \\
40  & EW\_FE\_MGII	          &  DOUBLE	&  Rest frame equivalent width of Fe within 2200-3090\AA in \AA \\
%43	& EW\_FE\_MGII\_ERR	      &  DOUBLE	&  Uncertainty in EW\_Fe\_MgII in \AA \\
41	& LINE\_NPIX\_MGII        &	 LONG	&  Number of good pixels for the restframe 2700-2900 \AA  \\
42	& LINE\_MED\_SN\_MGII	  &  DOUBLE	&  Median S/N per pixel for the restframe 2700-2900 \AA   \\
%46	& LINE\_REDCHI2\_MGII	  &  DOUBLE	&  Reduced $\chi^{2}$ for the MgII fit; -1 if not fitted   \\
43  & LINE\_FLAG\_MGII        &  LONG   &  Flag for the quality in MgII fitting \\
\hline
44	& FWHM\_CIV	              &  DOUBLE &	FWHM of the whole CIV in \kmps \\
%69	& FWHM\_CIV\_ERR	      &  DOUBLE	&   Uncertainty in FWHM\_CIV in \kmps \\
45	& EW\_CIV	              &  DOUBLE	&   Rest frame equivalent width of the whole CIV in \AA \\
%71	& EW\_CIV\_ERR	          &  DOUBLE	&   Uncertainty in EW\_CIV in \AA \\
46	& FWHM\_BROAD\_CIV	      &  DOUBLE &	FWHM of the broad CIV in \kmps \\
%73	& FWHM\_BROAD\_CIV\_ERR	  &  DOUBLE	&   Uncertainty in FWHM\_BROAD\_CIV in \kmps \\
47	& EW\_BROAD\_CIV	      &  DOUBLE	&   Rest frame equivalent width of the broad CIV in \AA \\
%75	& EW\_BROAD\_CIV\_ERR	  &  DOUBLE	&   Uncertainty in EW\_BROAD\_CIV in \AA \\
48	& FWHM\_NARROW\_CIV	      &  DOUBLE &	FWHM of the narrow CIV in \kmps \\
%77	& FWHM\_NARROW\_CIV\_ERR  &  DOUBLE	&   Uncertainty in FWHM\_NARROW\_CIV in \kmps \\
49	& EW\_NARROW\_CIV	      &  DOUBLE	&   Rest frame equivalent width of the narrow CIV in \AA \\
%79	& EW\_NARROW\_CIV\_ERR	  &  DOUBLE	&   Uncertainty in EW\_NARROW\_CIV in \AA \\
50  & LINE\_NPIX\_CIV	      &  LONG	&   Number of good pixels for the restframe 1500-1600 \AA  \\
51	& LINE\_MED\_SN\_CIV	  &  DOUBLE	&   Median S/N per pixel for the restframe 1500-1600 \AA   \\
%55	& LINE\_REDCHI2\_CIV	  &  DOUBLE	&   Reduced $\chi^{2}$ for the CIV fit; -1 if not fitted \\
52  & LINE\_FLAG\_CIV         &  LONG   &  Flag for the quality in CIV fitting \\
\hline
53    & ALPHA\_LAMBDA\_1   &  DOUBLE &   Wavelength power-law index from $\sim$ 1300 to 4661 \AA \\
54    & ALPHA\_LAMBDA\_2   &  DOUBLE &   Wavelength power-law index redward of 4661 \AA \\
55    & MODEL\_PHOT\_REDCHI2 &  DOUBLE  & Reduced chi-square  \\ 
56	& LOGL5100	              & DOUBLE	&   Monochromatic luminosity at 5100\AA (erg/s) \\
57	& LOGL3000	              & DOUBLE	&   Monochromatic luminosity at 3000\AA (erg/s) \\
58	& LOGL1350	              & DOUBLE	&   Monochromatic luminosity at 1350\AA (erg/s)  \\
59	& LOGBH\_HB               &	DOUBLE	&   Virial BH mass based on \hb\ (M$_{\sun}$) \\
60	& LOGBH\_MgII             &	DOUBLE	&   Virial BH mass based on MgII (M$_{\sun}$) \\
61	& LOGBH\_CIV              &	DOUBLE	&   Virial BH mass based on CIV(M$_{\sun}$) \\
\hline
62  & FPEAK                       & DOUBLE  &  FIRST peak flux density at 20 cm in mJy \\
63  & BAL\_FLAG                & STRING  &  Broad absorption trough flag \\
64  & FLAG\_DATA\_MINING & STRING &  Flag of quasar selections based on data mining \\
65  & FLAG\_OI\_COLOR      & STRING  &  Flag of quasar selections based on optical-infrared colour \\
66  & FLAG\_OTHERS          & STRING  & Flag of quasar selections from radio, X-ray, serendipitous algorithms. \\
\enddata
\end{deluxetable*}
%\end{center}

\clearpage
%%%%%%%%%%%%%%%%%%%
\begin{deluxetable*}{cccccccc}
\tabletypesize{\tiny}
\tablecaption{Parameters of the 2741 known quasars. \label{known_quasar}}
\tablewidth{0pt}
\tablehead{
\colhead{NAME} & \colhead{MJD}   & \colhead{PLANID}  & \colhead{SPID} & \colhead{FIBERID} & 
\colhead{FLAG\_DATA\_MINING}   & \colhead{FLAG\_OI\_COLOR} & \colhead{FLAG\_OTHERS}}
\startdata
000009.38+135618.4&56214&EG000024N121601B01&11&084&/XD/Richards09&/W/U&\\
000015.17+004833.3&56213&EG000023N024031B01&01&155&/XD/Richards09&/W/U&\\
000116.00+141123.0&56214&EG000024N121601B01&11&228&/SVM/Richards09&/W/U&\\
000234.51+051344.1&55893&F9302&14&194&&/W/U&\\
000338.63+114138.3&56214&EG000024N121601B01&08&011&/SVM/Richards09&/W/U&\\
000520.22+052410.7&55893&F9302&03&110&&/W/U&\\
000725.53+133313.5&56214&EG000024N121601B01&12&152&/Richards09&/W/U&\\
000727.05+024113.1&56213&EG000023N024031B01&13&092&/Richards09&/W/U&\\
004342.54+372519.9&56221&M31012N38B1&10&203&&/W&\\
004448.32+372114.7&55862&M6201&10&216&&/W&\\
004547.23+041023.4&55892&F9202&16&193&&/W/U&\\
004610.29+010644.1&55892&F9202&02&143&&/W&\\
004613.78+004410.3&55892&F9202&02&216&/SVM&/W&/Xray\\
004818.98+394111.7&55863&M31\_011N40\_B1&08&102&&/W&\\
005008.49+011330.1&55892&F9202&01&163&/SVM&&/Xray\\
005050.74+353642.9&55874&M31\_012N38\_B1&01&214&&/W&\\
\enddata
\tablecomments{Table 2 is published in its entirety in the electronic edition of the Astronomical Journal. A portion is shown here for guidance regarding its form and content.}
\end{deluxetable*}

\end{document}